\documentclass[11pt,english,floatfix,superscriptaddress,aps,prd,preprint,showkeys,nofootinbib]{revtex4}
\usepackage{amsmath}
\usepackage{amssymb}
\usepackage{amsbsy}
\usepackage{amsfonts}
\usepackage{amsopn}
\usepackage{amstext}
\usepackage{graphicx}
\usepackage[english]{babel}
\usepackage{color}
\usepackage{slashed}
\usepackage{esint}
\usepackage[dvips]{epsfig}
\usepackage[dvips]{graphicx}
\usepackage{float}
\usepackage{units}
\usepackage{textcomp}
\usepackage{wasysym}
\usepackage{hyperref}
\usepackage{slashed}

\begin{document}

\title{New mechanism for fermion localization in $f(T,T_G)$-brane}

\author{Allan R. P. Moreira}
\email{allan.moreira@fisica.ufc.br}
\affiliation{Research Center for Quantum Physics, Huzhou Normal University, Huzhou, 
313000, P. R. China.}
\affiliation{Secretaria da Educaç\~{a}o do Cear\'{a} (SEDUC), Coordenadoria 
Regional de Desenvolvimento da Educaç\~{a}o (CREDE 9),  Horizonte, Cear\'{a}, 
62880-384, Brazil.}
\author{Fernando M. Belchior}
\email{belchior@fisica.ufc.br}
\affiliation{Departamento de Física, Universidade Federal da Paraíba, Centro de Ciências Exatas e da Natureza, 58051-970, João Pessoa, Paraíba, Brazil}
\author{Guo-Hua Sun}
\email{sunghdb@yahoo.com}
\affiliation{Centro de Investigaci\'{o}n en Computaci\'{o}n, Instituto 
Polit\'{e}cnico Nacional, UPALM, CDMX 07700, M\'{e}xico}
\author{Shi-Hai Dong}
\email{dongsh2@yahoo.com}
\affiliation{Research Center for Quantum Physics, Huzhou Normal University, Huzhou, 
313000, P. R. China.}
\affiliation{Centro de Investigaci\'{o}n en Computaci\'{o}n, Instituto 
Polit\'{e}cnico Nacional, UPALM, CDMX 07700, M\'{e}xico}

\begin{abstract}
We investigate the localization of fermionic fields in a five-dimensional braneworld scenario within the framework of modified teleparallel gravity described by a general $f(T,T_G)$ function. Considering a non-minimal coupling between a Dirac spinor and the torsional invariants, we derive the effective Schr\"odinger-like equations governing the Kaluza--Klein modes. We showed that the contribution of the teleparallel Gauss-Bonnet term significantly modifies the effective potentials and, consequently, the localization properties. The zero-mode analysis reveals that only one chiral component can be localized on the brane, with the degree of confinement depending on the chosen model. In the massive sector, the spectrum is continuous, but resonant states arise due to the internal structure of the potentials. Additionally, we employ information-theoretic measures, such as Shannon entropy and relative probability, to quantify the localization mechanism. Our results show that the torsional modifications induce a nontrivial redistribution of information, exhibiting stronger localization. These findings highlight the role of higher-order torsional terms in shaping fermionic localization and resonance structures in braneworld scenarios.
\end{abstract}

\maketitle

\section{Introduction}

The idea of extending spacetime beyond four dimensions can be traced back to the seminal proposals of Kaluza \cite{Kaluza1921} and Klein \cite{Klein1926}, who originally aimed to unify Einstein's general relativity with Maxwellian electromagnetism within a higher-dimensional framework. Although these early constructions assumed compact extra dimensions, subsequent developments introduced scenarios in which the extra dimensions may be large or even non-compact \cite{Arkani-Hamed:1998jmv, rs, rs2, dsh}. This shift significantly expanded the phenomenological scope of higher-dimensional models. In particular, the Randall–Sundrum setup provided a fascinating mechanism for gravitational localization, giving rise to the braneworld paradigm. Parallel to this, alternative constructions such as the Gregory–Rubakov–Sibiryakov model \cite{Gregory:2000jc}, the universal extra dimension scenario \cite{Appelquist:2000nn}, and the Dvali–Gabadadze–Porrati framework \cite{Dvali:2000hr} further enriched the landscape of extra-dimensional theories. Additionally, considerable effort has been devoted to modeling smooth (thick) brane configurations generated by different types of matter fields, including scalar, vector, and fermionic sources \cite{DeWolfe:1999cp, Gremm1999, Csakil, Gremm2000, Dzhunushaliev:2009, Dzhunushaliev:2010, Herrera-Aguilar:2009, Dzhunushaliev:2007,  Goldberger1999, Bazeia2008, Geng:2015kvs, Dzhunushaliev:2011mm,1110,1111,1112,1113}.

From a different perspective, the need to go beyond general relativity is also motivated by both theoretical and observational challenges. At the fundamental level, the non-renormalizable character of GR suggests that it should be regarded as an effective theory, while at the cosmological scale, unresolved issues such as the cosmological constant problem and current observational tensions—including discrepancies in the Hubble parameter and matter clustering amplitude—provide further impetus for modified gravity approaches \cite{Abdalla:2022yfr}. In response, a wide class of extensions to GR has been developed \cite{CANTATA:2021ktz, Capozziello:2011et, Cai:2015emx}. A common strategy consists of generalizing the Einstein–Hilbert action, leading to theories such as $f(R)$ gravity \cite{Capozziello:2002rd, DeFelice:2010aj,2110,2111,2112,2113}, $f(G)$ models \cite{Nojiri:2005jg, DeFelice:2008wz}, cubic curvature corrections \cite{Asimakis:2022mbe}, and Lovelock gravity \cite{Lovelock:1971yv, Deruelle:1989fj}. Scalar-tensor formulations also play a central role, notably Horndeski theory \cite{Horndeski:1974wa} and its generalizations such as Galileon models \cite{DeFelice:2010nf, Deffayet:2011gz, Kobayashi:2010cm, DeFelice:2011bh,3110,3111,3112}.

An alternative geometric description is provided by the Teleparallel Equivalent of General Relativity, in which torsion, rather than curvature, encodes gravitational interactions. This framework admits several nontrivial generalizations, including $f(T)$ gravity \cite{ Chen:2010va,4110,4111,4112}, extensions involving the teleparallel Gauss–Bonnet term such as $f(T, T_G)$ theories \cite{Kofinas:2014owa, Kofinas:2014daa}, and models depending on both torsion and the boundary term $B$, namely $f(T, B)$ gravity \cite{Bahamonde:2015zma, Bahamonde:2016grb}. Motivated by analogies with scalar-tensor constructions \cite{Geng:2011aj}, one can further introduce scalar fields nonminimally coupled to torsional invariants, giving rise to scalar–torsion theories, including the teleparallel counterpart of Horndeski gravity \cite{Bahamonde:2019shr, Bahamonde:2020cfv, Capozziello:2023foy, Aldrovandi}. 

The connection between braneworld frameworks and modified gravitational dynamics has motivated a growing line of research in which extra-dimensional scenarios are constructed within extensions of general relativity. In this context, particular attention has been devoted to the localization properties of fermionic fields in teleparallel-based models. For instance, studies have investigated the role of Yukawa-type interactions in achieving fermion trapping in $f(T)$ gravity \cite{Yang2012} and in its generalized counterpart $f(T, B)$ \cite{Moreira20211}. Despite its widespread use, the standard Yukawa coupling may not fully capture how modifications in the underlying gravitational sector affect the localization mechanism of fermions on the brane. This limitation suggests the need for alternative approaches, such as the introduction of more general non-minimal couplings between fermionic fields and geometric quantities, which can provide a deeper understanding of the influence of spacetime structure on localization phenomena.

From this perspective, the realizations of gravity-based branes $f(T,T_G)$ have not yet been analyzed and may provide a particularly rich framework for exploring the combined effects of extra dimensions, torsion-driven dynamics, and higher-order geometric contributions. Unlike the four-dimensional case, the teleparallel Gauss–Bonnet term $T_G$ in five dimensions is dynamically nontrivial and actively contributes to the bulk field equations. As a consequence, thick-brane configurations in this setup can display features that are markedly distinct from those found in standard TEGR-based scenarios as well as in curvature-based Gauss–Bonnet models. A key aspect of these theories lies in the nontrivial dependence of the action on $f_{T_G}$ and its derivatives along the extra dimension, which introduces additional differential structures into the system. 
From a phenomenological point of view, such modifications can impact the localization properties of the fermion zero mode, reshaping the spectrum of massive Kaluza-Klein excitations to include possible resonant states.

This paper is organized as follows. In Sec.~\ref{s2}, we establish the theoretical framework of five-dimensional $f(T,T_G)$ gravity within the teleparallel formulation, deriving the corresponding field equations and introducing the warped brane configuration. We then investigate the localization of spin-$1/2$ fermions through a non-minimal coupling with the torsional invariants, obtaining the associated Schr\"odinger-like equations and analyzing both zero-mode and massive spectra for specific functional forms of $f(T,T_G)$. In Sec.~\ref{s3}, we complement the standard localization analysis by employing information-theoretic tools, including Shannon entropy measures, to characterize the spatial distribution and uncertainty of the fermionic modes, as well as the resonant structure of the massive sector via relative probability. Finally, in Sec.~\ref{s4}, we summarize our main results and discuss their physical implications.

\section{ Fermion localization $f(T,T_G)$ gravity}\label{s2}

In this section, we investigate the localization of fermionic fields in the context of five-dimensional $f(T,T_G)$ teleparallel gravity. We begin by establishing the geometric framework and deriving the corresponding field equations in terms of the torsional invariants $T$ and $T_G$. We then focus on the fermionic sector, considering a spin-$1/2$ field non-minimally coupled to the geometry through a function of $T$ and $T_G$. By performing a Kaluza--Klein decomposition, we derive the effective Schr\"odinger-like equations and analyze the resulting potentials. Finally, we examine the localization properties of both zero modes and massive modes, highlighting the role of the modified gravity function in shaping the fermionic spectrum.

\subsection{Geometric setup and field equations}

We formulate the five-dimensional theory in the teleparallel framework, where gravity is described by the fünfbein field $e^{a}{}_{M}$, with tangent-space indices $a=0,\dots,4$ and spacetime indices $M=0,\dots,4$. The spacetime metric is induced via \cite{Aldrovandi}
\begin{align}
g_{MN} = \eta_{ab}\, e^{a}{}_{M} e^{b}{}_{N},
\end{align}
with $\eta_{ab}=\mathrm{diag}(-,+,+,+,+)$ and $e=\det(e^{a}{}_{M})=\sqrt{-g}$. In this conjecture, instead of the Levi--Civita connection, we employ the curvature-free Weitzenb\"ock connection, namely
\begin{align}
\widetilde{\Gamma}^{P}{}_{MN} = e_{a}{}^{P}\partial_{N} e^{a}{}_{M},
\end{align}
which leads to a nonvanishing torsion tensor,
\begin{align}
T^{P}{}_{MN} = \widetilde{\Gamma}^{P}{}_{NM} - \widetilde{\Gamma}^{P}{}_{MN}.
\end{align}
The torsion scalar is constructed from irreducible quadratic combinations of $T^{P}{}_{MN}$ \cite{Aldrovandi}. Explicitly, such a scalar reads
\begin{align}
T = \frac{1}{4}T_{PMN}T^{PMN} + \frac{1}{2}T_{PMN}T^{NMP} - T_{P}T^{P},
\end{align}
where $T_{P}=T^{M}{}_{MP}$. Besides, since we interested in incorporating higher-order contributions, we need to introduce the teleparallel Gauss-Bonnet invariant $T_G$, built from the contortion tensor given by
\begin{align}
K^{P}{}_{MN} = \frac{1}{2}\left(T_{M}{}^{P}{}_{N} + T_{N}{}^{P}{}_{M} - T^{P}{}_{MN}\right).
\end{align}
In differential-form language, $T_G$ is constructed from quartic contractions of the contortion and its derivatives, yielding a scalar that differs from the standard Gauss-Bonnet invariant by a total divergence. This ensures that, unlike in four dimensions, $T_G$ contributes dynamically in $D=5$. With these definitions in hand, we can write the teleparallel Gauss-Bonnet as follows
\begin{align}\label{eq6}
T_G&=\frac{1}{(D-4)!}\,\epsilon_{a_1\cdots a_D}
\Big(
K^{a_1}{}_{c}\wedge K^{ca_2}\wedge K^{a_3}{}_{d}\wedge K^{da_4}
-2\,K^{a_1a_2}\wedge K^{a_3}{}_{c}\wedge K^{cd}\wedge K_{d}{}^{a_4}
\nonumber\\&+2\,K^{a_1a_2}\wedge D K^{a_3}{}_{c}\wedge K^{ca_4}
\Big)\wedge e^{a_5}\wedge\cdots\wedge e^{a_D},
\end{align}
where $K^{ab}$ is the contortion 1-form and $D$ denotes the exterior covariant derivative.  The fundamental identity is that the Levi-Civita Gauss-Bonnet scalar $\bar G$, which is constructed from the Levi--Civita connection, and $T_G$ differ by a total derivative, so that $\bar G=-T_G+\nabla_M(\cdots)$, generalizing the TEGR relation $\bar R=-T+\nabla_M(\cdots)$.

Moreover, the gravitational dynamics is governed by the action
\begin{align}
S = \int d^5x\, e \left[\frac{1}{4} f(T,T_G) + \mathcal{L}_m \right],
\end{align}
where $f(T,T_G)$ is an arbitrary function of the torsional invariants. Varying the action with respect to the fünfbein leads to modified field equations of the form
\begin{align}
\mathcal{E}_{ab} = 4\,\Theta_{ab},
\end{align}
where $\Theta_{ab}:=\Theta_a{}^{M}e_{bM}$ denotes the matter energy-momentum tensor and  \cite{KofinasSaridakis2014TTG}
\begin{align}\label{eq11}
\mathcal{E}_{ab}
&=
2\Big(H_{[ac]b}+H_{[ba]c}-H_{[cb]a}\Big)_{,}{}^{c}
+2\Big(H_{[ac]b}+H_{[ba]c}-H_{[cb]a}\Big)\,C^{d}{}_{dc}
\nonumber\\
&
+\Big(2H_{[ac]d}+H_{dca}\Big)\,C_{b}{}^{cd}
+4H_{[db]c}\,C_{a}{}^{dc}
+T_{acd}\,H^{cd}{}_{b}
-h_{ab}
+\Big(f-Tf_T-T_G f_{T_G}\Big)\eta_{ab},
\end{align} 
encodes the gravitational sector. The latter contains contributions from both $T$ and $T_G$, which can be organized through the decomposition \cite{KofinasSaridakis2014TTG}
\begin{align}
H^{abc} = f_T H^{abc}_{(T)} + f_{T_G} H^{abc}_{(G)},
\end{align}
with $f_T=\partial f/\partial T$ and $f_{T_G}=\partial f/\partial T_G$. The explicit expressions follow from functional differentiation of the invariants with respect to the tetrad variables.

\subsection{Warped brane configuration}

We can construct thick brane solutions by considering the following warped metric \cite{Moreira20211}
\begin{align}
ds^2 = e^{2A(y)} \eta_{\mu\nu} dx^\mu dx^\nu + dy^2,
\end{align}
which is supported by the diagonal fünfbein
\begin{align}
e^{a}{}_{M} = \mathrm{diag}(e^{A}, e^{A}, e^{A}, e^{A}, 1),
\end{align}
where $A(y)$ is the warp function responsible for generating a thick brane configuration. Within this background, the torsion scalar reduces to $T = -12 A'^2$, while the teleparallel Gauss-Bonnet invariant becomes
$T_G=24(5A'^4+4A'^2A'')$. Substituting the ansatz into the field equations yields two independent gravitational equations,
\begin{eqnarray}
 P(y)&=&\frac{f}{4} + 6A'^2 f_T - 24A'^2(A'^2 + A'')f_{T_G} + 24A'^3 f_{T_G}' 
 ,\nonumber\\
\rho(y)&=&\frac{f}{4} + 6A'^2 \Big(f_T +f_{T_G}''\Big)+ \frac{3}{2}\Big(A'' f_T + A' f_T'\Big)
 - 18A'^2 \Big(A'' f_{T_G} -A'f_{T_G}'\Big) \nonumber\\
&+&  12A'A'' f_{T_G}' -24A'^4 f_{T_G}.
\end{eqnarray}
being $f_T$ and $f_{T_G}$ depend implicitly on $y$ through $T$ and $T_G$, their derivatives can be expressed via the chain rule,
$f_T' = f_{TT} T' + f_{TG} T_G'$, 
$f_{T_G}' = f_{GT} T' + f_{GG} T_G'$,
with higher derivatives obtained analogously.

As a consistency check, in the limit $f(T,T_G)=-T$, the system reduces to the standard Einstein-scalar equations,
\begin{eqnarray}
 P(y)&=&3A'^2,\nonumber\\
\rho(y)&=&-\frac{3}{2}A'' - 3A'^2,
\end{eqnarray}
recovering the usual thick brane scenario.

\subsection{Fermionic sector and non-minimal geometric coupling}

We investigate the confinement of spin-$1/2$ fields in a five-dimensional warped geometry arising from a teleparallel modified gravity theory of the $f(T,T_G)$ type. The spacetime is assumed to possess a single extra spatial dimension, with the line element written in conformal coordinates as
\begin{equation}
ds^{2} = e^{2A(z)}\left(\eta_{\mu\nu}dx^\mu dx^\nu + dz^{2}\right).
\end{equation}
To model a smooth domain wall structure, we adopt a warp profile of the form \cite{Moreira20211}
\begin{equation}
A(z) = -p \ln\left[\mathrm{arcsinh}(\lambda z)\right],
\end{equation}
with $p$ controlling the deformation of the warp factor and $\lambda$ setting the characteristic thickness of the brane.

Beyond the gravitational sector, we introduce a Dirac spinor $\Psi$ propagating in the bulk, which interacts with the torsional geometry through a non-minimal coupling. The corresponding action is written as
\begin{equation}
\mathcal{S}_{1/2} = \int d^{5}x \, h \, \bar{\Psi}\left(\Gamma^{M}D_{M} - \xi f(T,T_G)\right)\Psi,
\end{equation}
where $h=\det(h^{a}{}_{M})$, $\Gamma^{M}=h_{a}{}^{M}\Gamma^{a}$ are the curved-space gamma matrices, and $D_{M}=\partial_{M}+\Omega_{M}$ denotes the spinorial covariant derivative. The spin connection is constructed from the contortion tensor as
\begin{equation}
\Omega_{M} = \frac{1}{4}K_{M}{}^{bc}\Gamma_{b}\Gamma_{c}.
\end{equation}

Adopting a chiral representation for the fermionic field, the five-dimensional Dirac equation reduces to
\begin{equation}
\left[\gamma^{\mu}\partial_{\mu} + \gamma^{4}\partial_{z} - \xi e^{A(z)} f(T,T_G)\right]\psi(x,z)=0.
\end{equation}

We perform the standard Kaluza--Klein decomposition
\begin{equation}
\psi(x,z) = \sum_{n}\left[\psi_{L,n}(x)\varphi_{L,n}(z) + \psi_{R,n}(x)\varphi_{R,n}(z)\right],
\end{equation}
where the four-dimensional spinors satisfy $\gamma^\mu\partial_\mu\psi_{L,R} = m \psi_{R,L}$ and $\gamma^{4}\psi_{L,R} = \mp \psi_{L,R}$.

This decomposition leads to the coupled first-order system
\begin{align}
\left[\partial_{z} + U(z)\right]\varphi_{L}(z) &= -m \varphi_{R}(z), \\
\left[\partial_{z} - U(z)\right]\varphi_{R}(z) &= m \varphi_{L}(z),
\end{align}
where the function
\begin{equation}
U(z) \equiv \xi e^{A(z)} f(T,T_G)
\end{equation}
acts as an effective superpotential.

Decoupling the system yields Schr\"odinger-like equations
\begin{align}
\left[-\partial_{z}^{2} + V_{L}(z)\right]\varphi_{L} &= m^{2}\varphi_{L}, \\
\left[-\partial_{z}^{2} + V_{R}(z)\right]\varphi_{R} &= m^{2}\varphi_{R},
\end{align}
with partner potentials given by
\begin{align}
V_{L}(z) &= U^{2}(z) - U'(z), \\
V_{R}(z) &= U^{2}(z) + U'(z).
\end{align}

In order to ensure physically acceptable localization properties, the function $f(T,T_G)$ must satisfy the following general requirements:
(i) it should display a nontrivial structure near the brane core, allowing for a transition around $z=0$;
(ii) it must approach a constant value asymptotically;
(iii) the effective superpotential $U(z)$ should vanish at large $|z|$, ensuring a well-defined vacuum behavior.

Motivated by these conditions, we consider the illustrative choices
\begin{align}
f_{1}(T,T_G) &= \sqrt{-T}+ \alpha T_G, \label{f1}\\
f_{2}(T,T_G) &= \sqrt{-T -\alpha T_G} \label{f2},
\end{align}
where $\alpha$ controls the relative contribution of the boundary term.

\begin{figure}[ht!]
\includegraphics[scale=0.64]{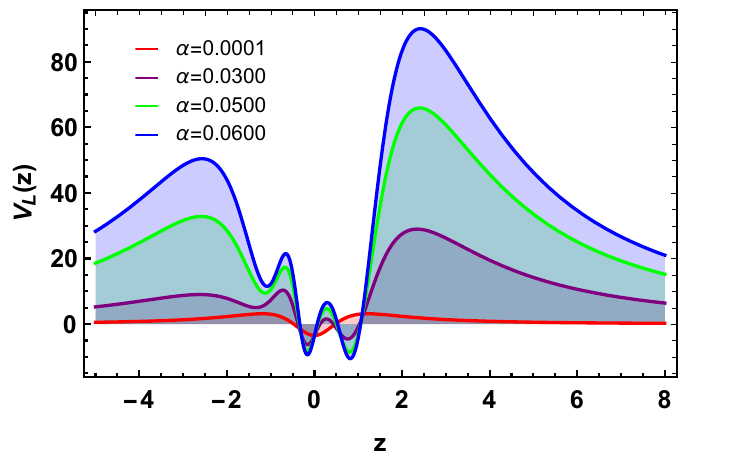} 
\includegraphics[scale=0.64]{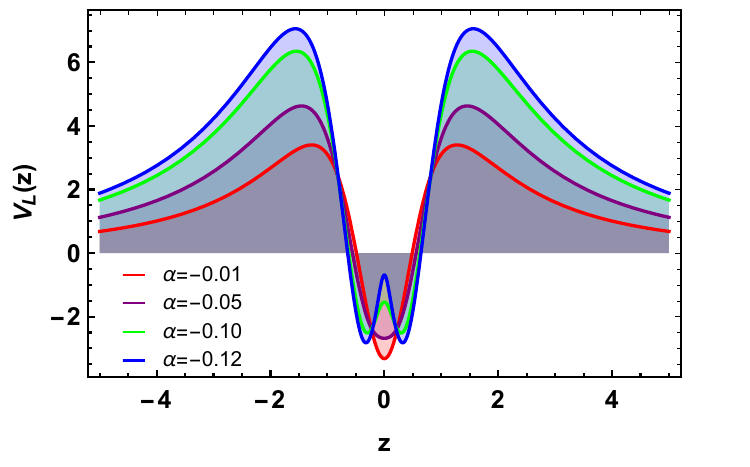}\\
(a) $f_{1}(T,T_G)$ defined in (\ref{f1}) \hspace{4cm}(b) $f_{2}(T,T_G)$ defined in (\ref{f2})\\
\caption{The effective potential for $\lambda=p=\xi=1$, and various 
values of the modified gravity parameter $\alpha$.
}
\label{fig1}
\end{figure}

Figure~\ref{fig1} displays the effective Schr\"odinger-like potentials associated with the left-chiral fermionic modes for the two functional choices of $f(T,T_G)$ given in Eqs.~(\ref{f1}) and (\ref{f2}). In panel (a), corresponding to $f_{1}(T,T_G)$, the potential exhibits a volcano-like structure, whose depth and asymmetry increase with the parameter $\alpha$. As $\alpha$ increases, the potential well deepens and becomes more distorted, indicating a stronger localization mechanism and the possible emergence of resonant states in the massive spectrum. In contrast, panel (b), associated with $f_{2}(T,T_G)$, shows a double-well profile with a central barrier around the brane core. This structure becomes more pronounced as $\alpha$ decreases, signaling the formation of quasi-bound states and a richer resonance pattern. In both cases, the asymptotic behavior of the potentials ensures the existence of a continuous spectrum of massive modes, while the modifications induced by $\alpha$ play a crucial role in shaping the localization properties and the internal structure of the fermionic spectrum.

\subsubsection{Zero-mode localization}

The massless sector ($m=0$) admits analytical solutions of the form
\begin{equation}
\varphi_{L0,R0}(z) \propto \exp\left[\pm \int U(z)\,dz\right].
\end{equation}

Normalizability requires
\begin{equation}
\int_{-\infty}^{\infty} |\varphi(z)|^{2} dz < \infty.
\end{equation}

Using the asymptotic behavior $e^{A(z)} \sim 1/|z|$ for large $|z|$, one finds that only one chiral component (left-handed for $\xi>0$) can be localized. The convergence condition leads to the constraint
\begin{equation}
\xi > \frac{\lambda}{2c},
\end{equation}
where $c = f(T,T_G)|_{|z|\to\infty}$.

The zero-mode profiles reflect the structure of the potentials: symmetric localization for $f_{1}$ and displaced localization for $f_{2}$, with increasing asymmetry as $\alpha$ grows.

\begin{figure}
\includegraphics[scale=0.64]{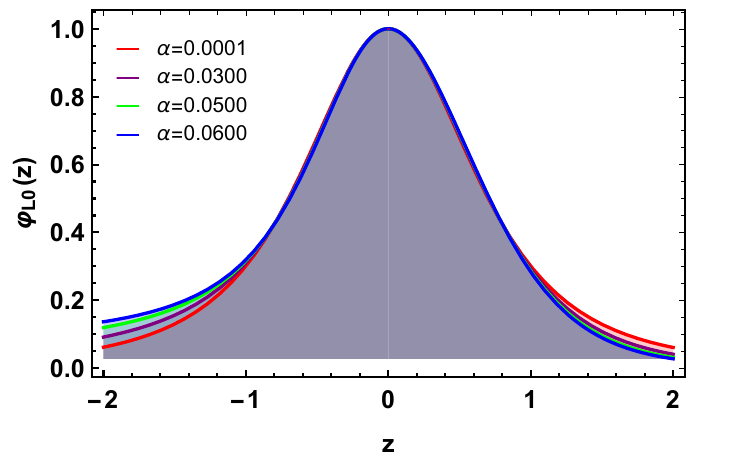} 
\includegraphics[scale=0.64]{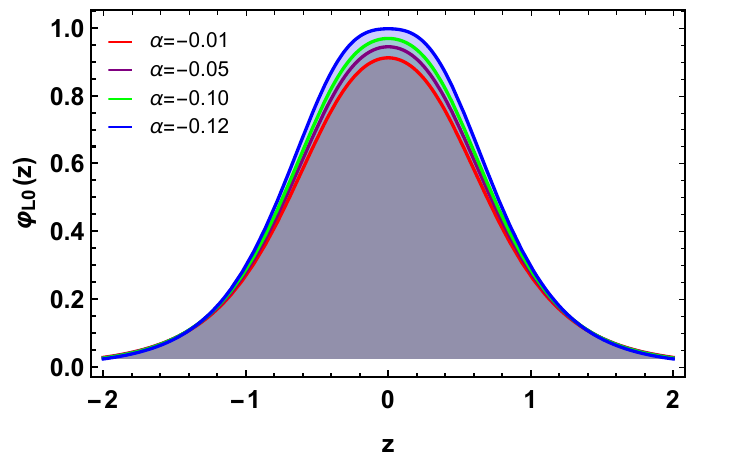}\\
(a) $f_{1}(T,T_G)$ defined in (\ref{f1}) \hspace{4cm}(b) $f_{2}(T,T_G)$ defined in (\ref{f2})\\
\caption{ The Zero-mode for $\lambda=p=\xi=1$, and various 
values of the modified gravity parameter $\alpha$.
\label{fig2}}
\end{figure}

Figure~\ref{fig2} illustrates the behavior of the normalized left-chiral fermionic zero-mode for the two choices of the function $f(T,T_G)$ given in Eqs.~(\ref{f1}) and (\ref{f2}), for different values of the parameter $\alpha$. In panel (a), corresponding to $f_1(T,T_G)$, the zero-mode exhibits an asymmetric profile located around the brane core at $z=0$. For values of $\alpha\rightarrow 0$, the symmetry is restored. As $\alpha$ increases, the amplitude near the origin becomes slightly larger, while the tails decay more rapidly, indicating a strengthening of the localization mechanism without altering the symmetry of the configuration. However, note that the left side decays to zero more slowly as the value of $\alpha$ increases. This behavior is related to the asymmetric behavior of the effective potential. In contrast, panel (b), associated with $f_{2}(T,T_G)$, also displays a localized zero-mode centered at the brane, but with a more pronounced sensitivity to negative values of $\alpha$. In this case, the peak becomes sharper and narrower as $|\alpha|$ increases, suggesting a higher degree of confinement. In both scenarios, the normalizability of the zero-mode is preserved, confirming that the modified torsional contributions encoded in $f(T,T_G)$ play a significant role in controlling the localization width and intensity of the fermionic ground state.

\subsubsection{Massive spectrum}

For $m \neq 0$, the equations must be solved numerically. Due to the parity properties of the potentials, the solutions split into even and odd modes satisfying \cite{Almeida2009,Liu2009,Liu2009a}
\begin{align}
\varphi_{\text{even}}'(0) &= 0, \quad \varphi_{\text{even}}(0) \neq 0, \\
\varphi_{\text{odd}}(0) &= 0, \quad \varphi_{\text{odd}}'(0) \neq 0.
\end{align}
The massive modes are not normalizable and correspond to continuum states. Their profiles resemble oscillatory waves in the bulk, although their amplitudes are modulated near the brane due to the potential structure.

\begin{figure}
\includegraphics[scale=0.66]{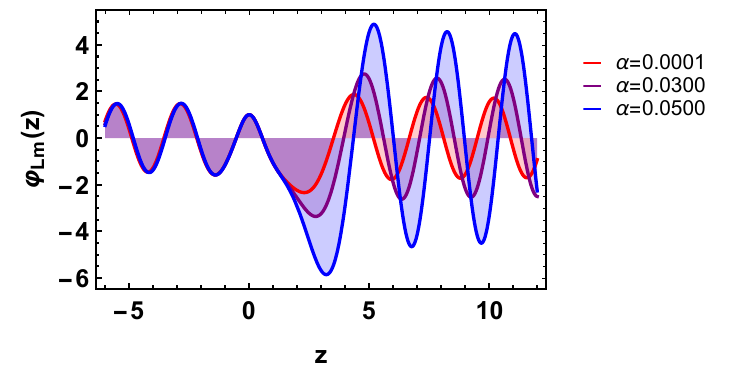} \hspace{-0.8cm}
\includegraphics[scale=0.66]{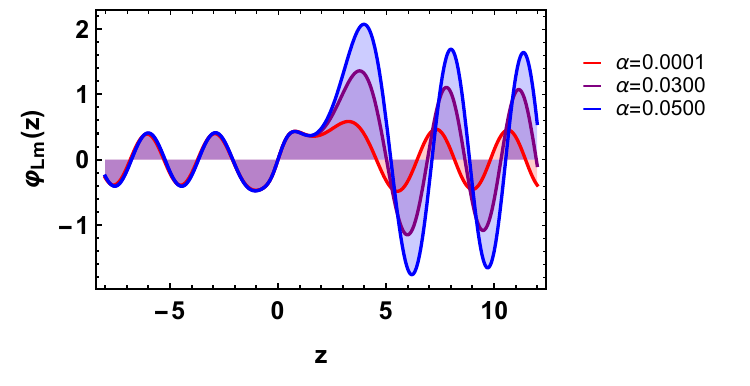}\\
(a) Even case \hspace{6cm}(b) Odd case\\
 \caption{
The massive modes for $\lambda=p=\xi=1$,  for the $f_1$ case of 
(\ref{f1}) and various values of the modified gravity parameter $\alpha$.
\label{fig3}}
\end{figure}

Figure~\ref{fig3} presents the profiles of the massive fermionic modes for the model defined by $f_{1}(T,T_G)$ in Eq.~(\ref{f1}), considering different values of the parameter $\alpha$. The solutions are separated according to their parity, with panel (a) corresponding to even modes and panel (b) to odd modes. In both cases, the wavefunctions exhibit an oscillatory behavior along the extra dimension, characteristic of non-localized continuum states. Near the brane core, however, the amplitudes are modulated by the structure of the effective potential, leading to a partial concentration of the modes. As $\alpha$ increases, the amplitude of the oscillations becomes more pronounced, indicating a stronger interaction between the fermionic modes and the modified gravitational background. The distinction between even and odd parity is reflected in the boundary conditions at the origin, with even modes attaining a nonvanishing value at $z=0$, while odd modes vanish at this point. Overall, the absence of normalizability confirms that these states belong to the continuous spectrum, although their enhanced amplitude near the brane suggests the possible emergence of resonant modes.

\begin{figure}
\includegraphics[scale=0.66]{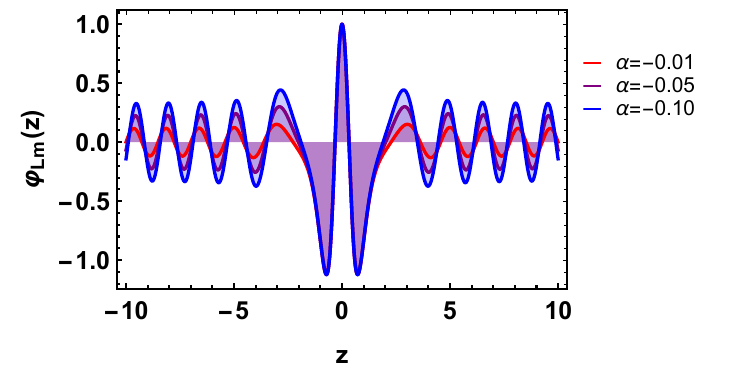} \hspace{-0.8cm}
\includegraphics[scale=0.66]{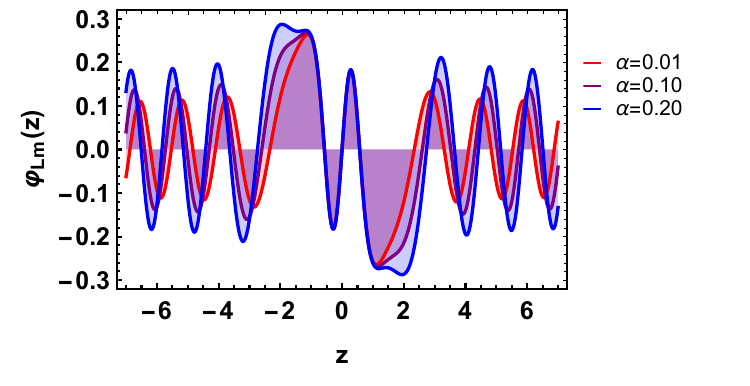} \\
(a) Even case \hspace{6cm}(b) Odd case\\
\caption{ 
The massive modes for $\lambda=p=\xi=1$, for the $f_2$ case of (\ref{f2})
and various values of the modified gravity parameter $\alpha$.
\label{fig4}}
\end{figure}

Figure~\ref{fig4} shows the profiles of the massive fermionic modes for the model defined by $f_{2}(T,T_G)$ in Eq.~(\ref{f2}), for different values of the parameter $\alpha$. As in the previous case, the solutions are separated into even and odd parity modes, displayed in panels (a) and (b), respectively. The wavefunctions exhibit the characteristic oscillatory behavior of continuum states, extending along the extra dimension without normalizability. However, in contrast to the $f_{1}(T,T_G)$ case, the presence of a more structured effective potential leads to a noticeable asymmetry and modulation of the amplitudes around the brane region. As $|\alpha|$ increases, the oscillations become more pronounced and exhibit localized enhancements near the brane core, suggesting a stronger interaction with the modified torsional background. In particular, the odd modes display a clear nodal structure at $z=0$, while the even modes remain finite at the origin, reflecting the imposed boundary conditions. The overall behavior indicates that the $f_{2}$ model supports a richer resonant structure, with an increased likelihood of quasi-localized massive states compared to the $f_{1}$ scenario.

\section{Information-theoretic characterization}\label{s3}

In this section, we employ information-theoretic tools to quantitatively assess the localization properties of fermionic modes in the braneworld setup. These measures provide complementary insight beyond the standard potential analysis, allowing one to characterize how the underlying geometry influences the spatial distribution and uncertainty of the fermionic states. For the zero modes, we focus on Shannon entropy, while for the massive sector, we analyze resonant behavior through a relative probability approach.

\subsection{Shannon entropy analysis}

The Shannon entropy offers a global measure of uncertainty associated with a probability distribution \cite{Shannon}. In the present context, it is constructed from the normalized fermionic density $|\varphi_{L0,R0}(z)|^{2}$, which encodes the spatial localization of the zero modes.

To access the complementary description in momentum space, we first introduce the Fourier transform of the probability density,
\begin{equation}
|\varphi_{L0,R0}(p_z)|^{2} = \frac{1}{\sqrt{2\pi}} \int_{-\infty}^{\infty} |\varphi_{L0,R0}(z)|^{2} e^{-ip_z z}\,dz.
\end{equation}

The Shannon entropies in position and momentum space are then defined as \cite{Beckner,Bialy}
\begin{align}
S_{z} &= -\int_{-\infty}^{\infty} |\varphi(z)|^{2}\ln |\varphi(z)|^{2}\,dz, \\
S_{p_z} &= -\int_{-\infty}^{\infty} |\varphi(p_z)|^{2}\ln |\varphi(p_z)|^{2}\,dp_z.
\end{align}
Furthermore, we can define the entropy densities of the system:
\begin{eqnarray}
\rho_s(z)&=&\vert\varphi_{L0, R0}(z)\vert^{2}\ln\vert\varphi_{L0, R0}(z)\vert^{2}, 
\\
\rho_s(p_z)&=&\vert\varphi_{L0, R0}(p_z)\vert^{2}\ln\vert\varphi_{L0, R0}
(p_z)\vert^{2}. 
\end{eqnarray}

These quantities satisfy the entropic uncertainty relation
\begin{equation}
S_{z} + S_{p_z} \geq 1 + \ln\pi,
\end{equation}
which provides a stronger formulation of uncertainty compared to the standard Heisenberg relation in one dimension. In our model, only the extra dimension experiences entropic changes within the system, and therefore $D=1$, i.e., $S_{z} + S_{p_z} \geq 2.14473$.

\begin{table}[ht!]
\begin{tabular}{|c||c|c|c|c|}
\hline
 & $\alpha$ & $S_{z}$ & $S_{p_z}$ & $S_{z}+S_{p_z}$\\ \hline
\hline
$f_1$  & 0.0001  & 0.70402 & 1.49104  & 2.19506  \\
       & 0.0010  & 0.70458 & 1.49084  & 2.19542  \\
       & 0.0050  & 0.72059 & 1.48516  & 2.20575\\ 
       & 0.0080  & 0.77127 & 1.46984  & 2.24111  \\ \hline \hline

$f_2$  & -0.001  & 0.65399 & 1.49104  & 2.14503   \\
       & -0.010  & 0.65866 & 1.49026  & 2.14893   \\
       & -0.050  & 0.69048 & 1.47604  & 2.16652 \\ 
       & -0.060  & 0.70250 & 1.47077  & 2.17327  \\ \hline 
\end{tabular}
\caption{Shannon entropy values in position and momentum space, for the two 
cases of the $f(T,T_G)$ function,   for  
$\xi=p=\lambda=1$.\label{tab1}}
\end{table}

Table~\ref{tab1} summarizes the numerical values of the Shannon entropies in position space ($S_{z}$) and momentum space ($S_{p_z}$), as well as their sum, for both functional forms $f_{1}(T,T_G)$ and $f_{2}(T,T_G)$, considering different values of the parameter $\alpha$. One observes that, for both models, an increase in $|\alpha|$ leads to a growth in $S_{z}$ accompanied by a slight decrease in $S_{p_z}$, reflecting a redistribution of information between position and momentum spaces. Physically, this behavior indicates that the fermionic zero-mode becomes less localized in the extra dimension as the torsional modification parameter increases, while simultaneously becoming more localized in momentum space. Despite these variations, the sum $S_{z}+S_{p_z}$ remains above the entropic uncertainty bound, confirming the consistency of the results with the fundamental uncertainty principle. Moreover, the $f_{2}(T,T_G)$ case systematically yields lower total entropy values compared to $f_{1}(T,T_G)$, suggesting a comparatively higher degree of localization of the fermionic modes. These results highlight the sensitivity of the information-theoretic measures to the underlying modified gravity structure, providing a complementary perspective on the localization mechanism.

\begin{figure*} [ht!]
\includegraphics[scale=0.65]{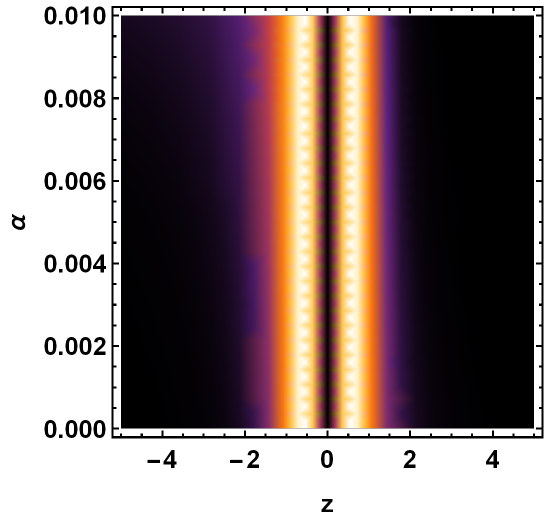} 
\includegraphics[scale=0.65]{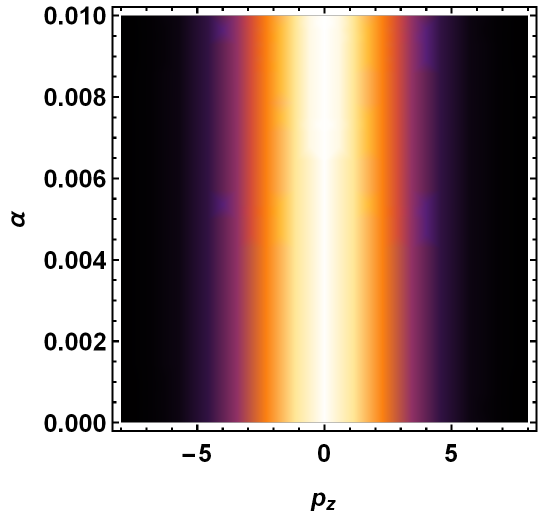}\\ 
(a) $\rho_s(z)$ for $f_{1}$ \hspace{3.5cm}(b) $\rho_s(p_z)$ for $f_{1}$ \\
\includegraphics[scale=0.65]{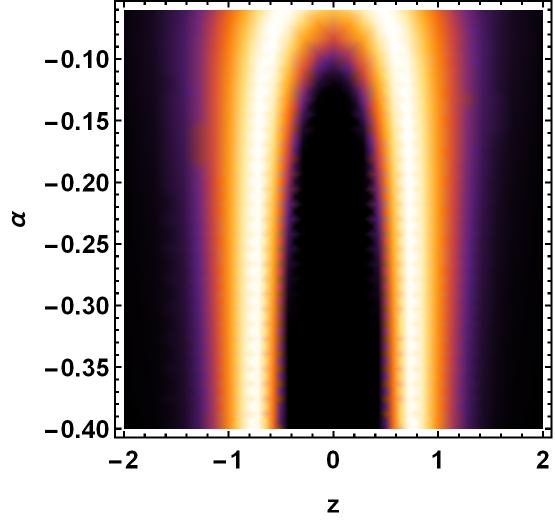} 
\includegraphics[scale=0.65]{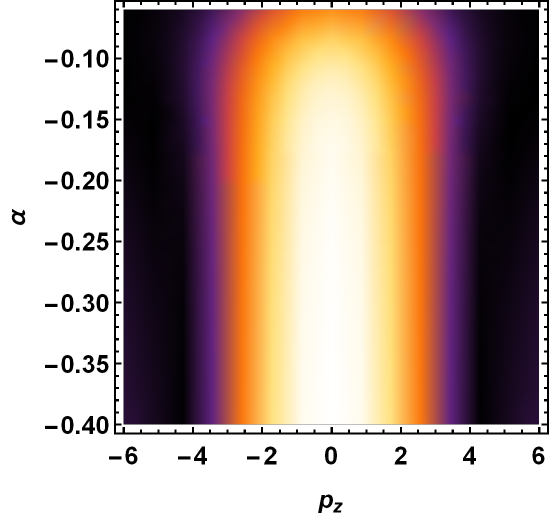} \\
(c) $\rho_s(z)$ for $f_{2}$  \hspace{3.5cm}(d) $\rho_s(p_z)$ for $f_{2}$ \\
\caption{ The entropic densities for $\lambda=p=\xi=1$.  
\label{fig6}}
\end{figure*}

Figure~\ref{fig6} displays the entropic density distributions $\rho_s(z)$ and $\rho_s(p_z)$ for the models $f_{1}(T,T_G)$ and $f_{2}(T,T_G)$, represented as density plots in the $(z,\alpha)$ and $(p_z,\alpha)$ planes. In these plots, the color scale encodes the magnitude of the entropic density, where the reddish–yellow regions correspond to higher values, indicating the dominant contributions to the entropy, while darker regions represent suppressed contributions. For the $f_{1}(T,T_G)$ case, panels (a) and (b) show a symmetric concentration of entropy around the brane core ($z=0$) and low momenta, with a gradual broadening as $\alpha$ increases, reflecting a redistribution of the fermionic information content. In contrast, the $f_{2}(T,T_G)$ case, depicted in panels (c) and (d), exhibits a more pronounced deformation of the density profiles, with the emergence of a central depletion in position space and a corresponding spreading in momentum space as $\alpha$ becomes more negative. This behavior signals a stronger sensitivity of the entropic structure to the modified torsional coupling. Overall, these density plots provide a clear visualization of how the parameter $\alpha$ controls the localization and uncertainty of the fermionic zero modes, with the high-density (reddish–yellow) regions identifying the most relevant domains contributing to the entropy of the system.

\subsection{Resonant structure and relative probability}

We now turn to the analysis of the massive fermionic spectrum. Although these modes are not strictly localized, certain states may exhibit enhanced amplitudes near the brane, behaving as resonances.

Such resonant modes arise when the effective potential forms a quasi-bound structure capable of temporarily trapping modes with $m^{2}$ below the height of the potential barrier. To quantify this effect, we introduce the relative probability \cite{Liu2009,Liu2009a}
\begin{equation}
P_{L,R}(m) = \frac{\int_{-z_b}^{z_b} |\varphi_{L,R}(z)|^{2}\,dz}{\int_{-z_{\text{max}}}^{z_{\text{max}}} |\varphi_{L,R}(z)|^{2}\,dz},
\end{equation}
which measures the fraction of the wavefunction localized within a finite region around the brane.

\begin{figure}
 \includegraphics[scale=0.66]{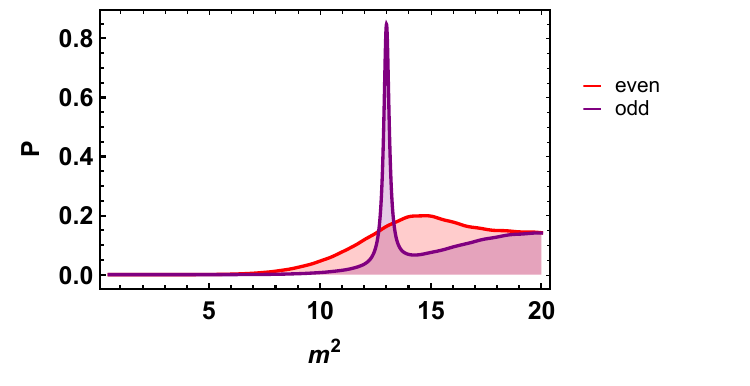}\hspace{-0.8cm}
\includegraphics[scale=0.66]{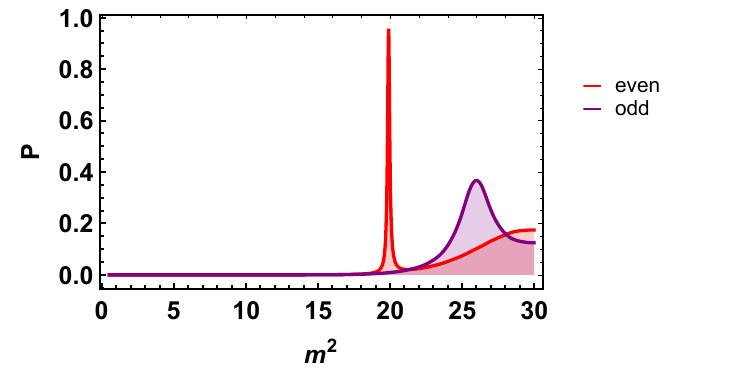}\\
(a) $f_{1}(T,T_G)$ defined in (\ref{f1}) \hspace{3.5cm}(b) $f_{2}(T,T_G)$ defined in (\ref{f2})\\
\caption{ The relative probability for $\xi=p=\lambda=1$ 
and $\alpha=0.1$
\label{fig5}}
\end{figure}

Figure~\ref{fig5} depicts the relative probability $P(m)$ as a function of the squared mass $m^{2}$ for both models $f_{1}(T,T_G)$ and $f_{2}(T,T_G)$, highlighting the resonant structure of the massive fermionic spectrum. In panel (a), corresponding to $f_{1}(T,T_G)$, the even and odd modes exhibit smooth distributions with a pronounced peak for the odd modes, indicating the presence of a resonant state with enhanced localization near the brane. The even modes, in contrast, display a broader profile with lower amplitude, suggesting weaker confinement. In panel (b), associated with $f_{2}(T,T_G)$, the resonance structure becomes more distinct, with sharper and more localized peaks, particularly for the even modes. The appearance of narrow and high-amplitude peaks signals the existence of quasi-bound states with longer lifetimes, which remain temporarily trapped in the vicinity of the brane before leaking into the bulk. Additionally, the shift in the position of the peaks between the two models reflects the influence of the modified torsional coupling on the mass spectrum. Overall, these results demonstrate that the choice of the function $f(T,T_G)$ plays a crucial role in determining the resonance pattern and the efficiency of fermion trapping in the extra dimension.

\section{Conclusions}\label{s4}

In this work, we analyzed fermion localization in a five-dimensional braneworld within $f(T,T_G)$ teleparallel gravity, considering a non-minimal coupling between the spin-$1/2$ field and the torsional invariants. From the resulting Schr\"odinger-like equations, we showed that the parameter $\alpha$ plays a key role in shaping the effective potentials and, consequently, the localization properties.
For the zero mode, only one chiral component can be localized, with the degree of confinement strongly dependent on the chosen function $f(T,T_G)$. The $f_{1}$ model leads to a volcano-like potential and weaker localization effects, while $f_{2}$ generates a more structured potential, enhancing confinement and asymmetry. In the massive sector, the spectrum is continuous, but resonant states emerge due to the internal structure of the potentials, being more pronounced in the $f_{2}$ case.

The information-theoretic analysis complements these results, showing that the parameter $\alpha$ controls the redistribution of localization in position and momentum spaces. In particular, the $f_{2}$ model exhibits lower total entropy, indicating stronger localization. The relative probability further confirms the presence of quasi-localized massive modes.
These results highlight the relevance of higher-order torsional terms in controlling fermion localization in braneworld scenarios. As future directions, one may consider more general forms of $f(T,T_G)$, include additional interactions, and extend the analysis to other fields or quantum information measures.

\subsection*{Acknowledgments} 
One of the authors, S.H. Dong, started this work during a research stay in China with permission from IPN, Mexico.  A.R.P. Moreira is grateful for the hospitality and support provided by the Research Center for Quantum Physics at Huzhou University. S.H. Dong and G.H. Sun acknowledge the partial support of the projects 20251087-SIP-IPN and 20251109-SIP-IPN, Mexico. F. M. Belchior would like to   express 
gratitude to the Conselho Nacional de Desenvolvimento Cient\'{i}fico e 
Tecnol\'{o}gico CNPq for grant No. 151845/2025-5.


\begin{thebibliography}{99}

\bibitem{Kaluza1921}
T.~Kaluza, 
Sitzungsber. Preuss. Akad. Wiss. Berlin (Math. Phys. ) \textbf{1921}, 966-972 
(1921).

\bibitem{Klein1926}
O.~Klein, 
Nature \textbf{118}, 516 (1926).

\bibitem{Arkani-Hamed:1998jmv}
N.~Arkani-Hamed, S.~Dimopoulos and G.~R.~Dvali, 
Phys. Lett. B \textbf{429}, 263-272 (1998).

\bibitem{rs}
    L.~Randall and R.~Sundrum, Phys.\ Rev.\ Lett.\  {\bf 83}, 4690 (1999). 
        
\bibitem{rs2}
    L.~Randall and R.~Sundrum, Phys.\ Rev.\ Lett.\  {\bf 83}, 3370 (1999).
\bibitem{dsh}
S.~H. Dong, Wave equations in higher dimensions, Springer, Dordrecht, Heidelberg, London, New York, 2011.


\bibitem{Gregory:2000jc}
R.~Gregory, V.~A.~Rubakov and S.~M.~Sibiryakov, 
Phys. Rev. Lett. \textbf{84}, 5928-5931 (2000).

\bibitem{Appelquist:2000nn}
T.~Appelquist, H.~C.~Cheng and B.~A.~Dobrescu, 
Phys. Rev. D \textbf{64}, 035002 (2001).

\bibitem{Dvali:2000hr}
G.~R.~Dvali, G.~Gabadadze and M.~Porrati, 
Phys. Lett. B \textbf{485}, 208-214 (2000).

\bibitem{DeWolfe:1999cp}
O.~DeWolfe, D.~Z.~Freedman, S.~S.~Gubser and A.~Karch, 
Phys. Rev. D \textbf{62}, 046008 (2000).

\bibitem{Gremm1999}
M.~Gremm, 
Phys. Lett. B \textbf{478}, 434 (2000).

\bibitem{Csakil}
C.~Csaki, J.~Erlich, T.~J.~Hollowood and Y.~Shirman, 
Nucl. Phys. B \textbf{581}, 309-338 (2000).

\bibitem{Gremm2000}
M.~Gremm, 
Phys. Rev. D \textbf{62}, 044017 (2000).

\bibitem{Dzhunushaliev:2009}
V.~Dzhunushaliev, V.~Folomeev and M.~Minamitsuji, 
Rept. Prog. Phys. \textbf{73}, 066901 (2010).

\bibitem{Dzhunushaliev:2010}
V.~Dzhunushaliev and V.~Folomeev, 
Gen. Rel. Grav. \textbf{43}, 1253-1261 (2011).

\bibitem{Herrera-Aguilar:2009}
A.~Herrera-Aguilar, D.~Malagon-Morejon, R.~R.~Mora-Luna and U.~Nucamendi, 
gap, ''
Mod. Phys. Lett. A \textbf{25}, 2089-2097 (2010).

\bibitem{Dzhunushaliev:2007}
V.~Dzhunushaliev, V.~Folomeev, D.~Singleton and S.~Aguilar-Rudametkin, 
Phys. Rev. D \textbf{77}, 044006 (2008).


\bibitem{Goldberger1999}
W.~D.~Goldberger and M.~B.~Wise, 
Phys. Rev. Lett. \textbf{83} (1999), 4922.

\bibitem{Bazeia2008}
D.~Bazeia, A.~R.~Gomes, L.~Losano and R.~Menezes, 
Phys. Lett. B \textbf{671} (2009), 402.

\bibitem{Geng:2015kvs}
W.~J.~Geng and H.~Lu, 
Phys. Rev. D \textbf{93}, no.4, 044035 (2016).

\bibitem{Dzhunushaliev:2011mm}
V.~Dzhunushaliev and V.~Folomeev, 
Gen. Rel. Grav. \textbf{44}, 253-261 (2012).


\bibitem{1110}
H.~L.~Jia, W.~D.~Guo, Y.~X.~Liu and Q.~Tan,
JHEP \textbf{06}, 117 (2025).

\bibitem{1111}
W.~Deng, S.~Long, Q.~Tan, Z.~C.~Chen and J.~Jing,
JHEP \textbf{01}, 066 (2026).

\bibitem{1112}
D.~Bazeia, A.~S.~Lob{\~a}o and M.~A.~Marques,
Nucl. Phys. B \textbf{1007}, 116662 (2024).


\bibitem{1113}
J.~L.~Rosa, A.~S.~Lob{\~a}o and D.~Bazeia,
Eur. Phys. J. C \textbf{82}, no.3, 191 (2022).


\bibitem{Abdalla:2022yfr}
E.~Abdalla, G.~Franco Abell\'an, A.~Aboubrahim, A.~Agnello, O.~Akarsu, 
Y.~Akrami, G.~Alestas, D.~Aloni, L.~Amendola and L.~A.~Anchordoqui, \textit{et 
al.}
JHEAp \textbf{34}, 49-211 (2022).

\bibitem{CANTATA:2021ktz}
E.~N.~Saridakis \textit{et al.} [CANTATA], 
Spinger (2021), 
[arXiv:2105.12582 [gr-qc]].
   
 

\bibitem{Capozziello:2011et}
  S.~Capozziello and M.~De Laurentis, 
  Phys.\ Rept.\  {\bf 509}, 167 (2011).
 

\bibitem{Cai:2015emx} 
  Y.~F.~Cai, S.~Capozziello, M.~De Laurentis and E.~N.~Saridakis, 
  Rept.\ Prog.\ Phys.\  {\bf 79}, 106901 (2016).
 

  
\bibitem{DeFelice:2010aj}
A.~De Felice and S.~Tsujikawa, 
Living Rev. Rel. \textbf{13}, 3 (2010).

\bibitem{Capozziello:2002rd}
S.~Capozziello, 
Int. J. Mod. Phys. D \textbf{11}, 483-492 (2002).

\bibitem{2110}
G.~G.~L.~Nashed and A.~Eid,
Eur. Phys. J. C \textbf{86}, no.3, 316 (2026).

\bibitem{2111}
G.~Montani, L.~A.~Escamilla, N.~Carlevaro and E.~Di Valentino,
Phys. Rev. D \textbf{113}, no.2, 023507 (2026).

\bibitem{2112}
A.~Valletta, G.~Montani, M.~G.~Dainotti and E.~Fazzari,
JHEAp \textbf{53}, 100612 (2026).


\bibitem{2113}
S.~Kibaro{\u{g}}lu,
Phys. Dark Univ. \textbf{47}, 101784 (2025).
 
\bibitem{Nojiri:2005jg}
  S.~Nojiri and S.~D.~Odintsov, 
  Phys.\ Lett.\ B {\bf 631}, 1 (2005).

\bibitem{DeFelice:2008wz}
A.~De Felice and S.~Tsujikawa, 
Phys. Lett. B \textbf{675}, 1-8 (2009).

 

\bibitem{Asimakis:2022mbe}
P.~Asimakis, S.~Basilakos and E.~N.~Saridakis, 
Eur. Phys. J. C \textbf{84}, no.2, 207 (2024).
   
\bibitem{Lovelock:1971yv}
  D.~Lovelock, 
  J.\ Math.\ Phys.\  {\bf 12}, 498 (1971).
  
\bibitem{Deruelle:1989fj}
N.~Deruelle and L.~Farina-Busto, 
Phys. Rev. D \textbf{41}, 3696 (1990).%
 
   
 
 
\bibitem{Horndeski:1974wa}
G.~W.~Horndeski, 
Int. J. Theor. Phys. \textbf{10}, 363-384 (1974).
 

\bibitem{DeFelice:2010nf}
A.~De Felice and S.~Tsujikawa, 
Phys. Rev. D \textbf{84}, 124029 (2011).


\bibitem{Deffayet:2011gz}
C.~Deffayet, X.~Gao, D.~A.~Steer and G.~Zahariade, 
Phys. Rev. D \textbf{84}, 064039 (2011).


\bibitem{Kobayashi:2010cm}
T.~Kobayashi, M.~Yamaguchi and J.~Yokoyama, 
Phys. Rev. Lett. \textbf{105}, 231302 (2010).

\bibitem{DeFelice:2011bh}
A.~De Felice and S.~Tsujikawa, 
JCAP \textbf{02}, 007 (2012).

\bibitem{3110}
W.~J.~Wolf, P.~G.~Ferreira and C.~Garc{\'\i}a-Garc{\'\i}a,
Phys. Rev. D \textbf{113}, no.2, 023551 (2026).

\bibitem{3111}
A.~Garoffolo, K.~Hinterbichler and M.~Trodden,
JHEP \textbf{09}, 115 (2025).

\bibitem{3112}
G.~Ye and A.~Silvestri,
Phys. Rev. D \textbf{111}, no.2, 023502 (2025).


\bibitem{Chen:2010va}
S.~H.~Chen, J.~B.~Dent, S.~Dutta and E.~N.~Saridakis, 
Phys. Rev. D \textbf{83}, 023508 (2011).


\bibitem{4110}
F.~Bajardi, D.~Blixt and S.~Capozziello,
Phys. Rev. D \textbf{111}, no.8, 084012 (2025).

\bibitem{4111}
C.~Wu, X.~Ren, Y.~Yang, Y.~M.~Hu and E.~N.~Saridakis,
Eur. Phys. J. C \textbf{85}, no.10, 1099 (2025).

\bibitem{4112}
O.~Akarsu, B.~Bulduk, A.~De Felice, N.~Kat{\i}rc{\i} and N.~M.~Uzun,
Phys. Rev. D \textbf{112}, no.8, 083532 (2025).


\bibitem{Kofinas:2014owa}
G.~Kofinas and E.~N.~Saridakis, 
Phys. Rev. D \textbf{90}, 084044 (2014).

\bibitem{Kofinas:2014daa}
G.~Kofinas and E.~N.~Saridakis, 
Phys. Rev. D \textbf{90}, 084045 (2014).

\bibitem{Bahamonde:2015zma}
S.~Bahamonde, C.~G.~B\"ohmer and M.~Wright, 
Phys. Rev. D \textbf{92}, no.10, 104042 (2015).

\bibitem{Bahamonde:2016grb}
S.~Bahamonde and S.~Capozziello, 
Eur. Phys. J. C \textbf{77}, no.2, 107 (2017).
 
\bibitem{Geng:2011aj}
C.~Q.~Geng, C.~C.~Lee, E.~N.~Saridakis and Y.~P.~Wu, 
Phys. Lett. B \textbf{704}, 384-387 (2011).
 
\bibitem{Bahamonde:2019shr}
S.~Bahamonde, K.~F.~Dialektopoulos and J.~Levi Said, 
Phys. Rev. D \textbf{100}, no.6, 064018 (2019).
 
 
\bibitem{Bahamonde:2020cfv}
S.~Bahamonde, K.~F.~Dialektopoulos, M.~Hohmann and J.~Levi Said, 
Class. Quant. Grav. \textbf{38}, no.2, 025006 (2020).

\bibitem{Capozziello:2023foy}
S.~Capozziello, M.~Caruana, J.~Levi Said and J.~Sultana, 
JCAP \textbf{03}, 060 (2023).


\bibitem{Aldrovandi}
R. Aldrovandi and J. G. Pereira, \textit{Teleparallel Gravity: An Introduction}, 
(Springer, Berlin, 2013).
 

\bibitem{Yang2012}
J.~Yang, Y.~-L.~Li, Y.~Zhong and Y.~Li, 
  Phys.\ Rev.\ D {\bf 85}, 084033 (2012).    

\bibitem{Moreira20211}
A.~R.~P.~Moreira, J.~E.~G.~Silva and C.~A.~S.~Almeida, 
Eur. Phys. J. C \textbf{81}, no.4, 298 (2021).

\bibitem{KofinasSaridakis2014TTG}
G.~Kofinas and E.~N.~Saridakis,
Phys.\ Rev.\ D {\bf 90}  (2014), 084044.


\bibitem{Almeida2009}
C.~A.~S.~Almeida, M.~M.~Ferreira, Jr., A.~R.~Gomes and R.~Casana, 
Phys. Rev. D \textbf{79}, 125022 (2009).

\bibitem{Liu2009}
Y.~X.~Liu, J.~Yang, Z.~H.~Zhao, C.~E.~Fu and Y.~S.~Duan, 
Phys. Rev. D \textbf{80}, 065019 (2009).

\bibitem{Liu2009a}
Y.~X.~Liu, H.~T.~Li, Z.~H.~Zhao, J.~X.~Li and J.~R.~Ren, 
JHEP \textbf{10}, 091 (2009).

\bibitem{Shannon}
C.~E.~Shannon, 
Bell Syst. Tech. J. \textbf{27}, no.3, 379-423 (1948).

\bibitem{Beckner}
W. Beckner, 
Ann. Math. {\bf 102}, 159 (1975).

\bibitem{Bialy}
I.~Bia\l{}ynicki-Birula and J.~Mycielski, 
Commun. Math. Phys. \textbf{44}, no.2, 129-132 (1975).






\end{thebibliography}
\end{document}